\documentclass[aps,prev,onecolumn,preprintnumbers,floatfix,nofootinbib]{revtex4-1}
\pdfoutput=1
\usepackage{geometry,amsmath,amsfonts}
\usepackage{slashed}
\usepackage{xcolor}
\usepackage{epsfig}
\usepackage{verbatim}
\usepackage{natbib}
\usepackage{braket}
\usepackage{adjustbox}
\usepackage{latexsym}
\usepackage{graphicx}
\usepackage[justification=RaggedRight]{caption}
\usepackage{amssymb}
\usepackage{subfig}
\usepackage{color}
\usepackage{multirow}
\usepackage{color}
\usepackage{rotating}
\usepackage{ifthen}
\usepackage{epsfig}

\makeatletter
\newcommand*\bigcdot{\mathpalette\bigcdot@{.5}}
\newcommand*\bigcdot@[2]{\mathbin{\vcenter{\hbox{\scalebox{#2}{$\m@th#1\bullet$}}}}}
\makeatother

\begin{document}
\vspace{1.5cm}

\title{Neutrino Mixing and Leptogenesis in a $L_e-L_\mu-L_\tau$ model}
\author{Giorgio Arcadi$^{a}$}
\email{giorgio.arcadi@unime.it}
\author{Simone Marciano$^{b,c}$}
\email{simone.marciano@uniroma3.it}
\author{Davide Meloni $^{b}$}
\email{davide.meloni@uniroma3.it}

\vspace{0.1cm}
 \affiliation{
${}^a$ 
 Dipartimento di Scienze Matematiche e Informatiche, Scienze Fisiche e Scienze della Terra, Universita degli Studi di Messina, Via Ferdinando Stagno d'Alcontres 31, I-98166 Messina, Italy
}

\vspace{0.1cm}
 \affiliation{
${}^b$ 
 Dipartimento di Matematica e Fisica, Universit\`a di Roma Tre, Via della Vasca Navale 84, 00146, Roma, Italy
}

\vspace{0.1cm}
 \affiliation{
${}^c$ 
INFN Sezione di Roma Tre, Via della Vasca Navale 84, 00146, Roma, Italy
}


\begin{abstract}
We present a simple extension of the Standard Model with three right-handed neutrinos in a SUSY framework, with an additional U(1$)_\text{F}$ abelian flavor symmetry with a non standard leptonic charge $L_e-L_\mu-L_\tau$ for lepton doublets and arbitrary right-handed charges. We show that the model is able to reproduce the experimental values of the mixing angles of the PMNS matrix and of the $r=\Delta m_\text{sun}^2/\Delta m_\text{atm}^2$ ratio, with only a moderate fine tuning of the Lagrangian free parameters.  The baryon asymmetry of the Universe is generated via thermal leptogenesis through CP-violating decays of the heavy right-handed neutrinos. We present a detailed numerical solution of the relevant Boltzmann equation accounting for the impact of the distribution of the asymmetry in the various lepton flavors. 

\end{abstract}
\maketitle

\newpage

\tableofcontents

\section{Introduction}
\label{sec:intro}

The Standard Model (SM) of particle physics has proven to be one of the most accurate theories to explain microscopic interactions at an unprecedented level. In spite of its many successes, it fails to account for relevant low energy data, such as the structure of fermion masses and mixings (in particular, the non-vanishing neutrino masses) and the value of the baryon asymmetry of the Universe (BAU), which is commonly expressed by the parameter:
\begin{equation}
  \eta_B\equiv \frac{n_B-n_{\bar{B}}}{n_\gamma}\bigg|_0\; ,  
\end{equation}
where $n_B,n_{\bar{B}}$ and $n_\gamma$ are the number densities of the baryons, antibaryons and photons, while the subscript "0" stands for "at present time". Latest observations provide a numerical value of $\eta_B\approx 6.1\cdot 10^{-10}$ \cite{Planck:2018vyg}.
In recent times, an enormous experimental progress has been made in our knowledge of the neutrino properties and it has been clearly shown that the lepton mixing matrix contains two large and one small mixing angle, and that the two independent mass-squared differences are both different from zero \cite{2020,deSalas:2020pgw,Capozzi:2021fjo}.
Although several abelian and non-abelian symmetries
acting on flavour space have been proposed to explain such a pattern, not an unique framework emerged as the optimal one \cite{Altarelli:2010gt}.
Thus, one is still motivated to explore scenarios where different symmetries and/or field (charge) assignments to the group representations are studied in details.  
In this context, a less explored possibility (compared to the most famous discrete non-abelian symmetries) is given by  the $U(1)_F$ flavor symmetry with non-standard leptonic charge $L_e - L_\mu - L_\tau$ for lepton doublets \cite{Petcov:1982ya} and arbitrary right-handed charges \cite{Altarelli:2004za}. 
As it is well known, in the limit of exact symmetry, the neutrino mass matrix assumes the following structure:
\begin{eqnarray} m_\nu =m_0
\left(
\begin{array}{ccc}
0& 1& x\\ 1& 0&0\\ x& 0&0
\end{array}
\right)\,,
\label{mass}
\end{eqnarray}
which leads to a spectrum of inverted type and to $\theta_{12} = \pi/4$, $\tan \theta_{23} = x$ (\emph{i.e.} large atmospheric mixing for $x\sim {\cal O}(1)$) and $\theta_{13} = 0$. While the previous texture can be considered a good Leading Order (LO) result, it evidently fails to reproduce two independent mass differences (two eigenvalues have the same absolute values) and, with the exception of the atmospheric angle, also fails in the correct description of the solar and reactor angles. Models based on the see-saw mechanism \cite{Lavoura:2000ci,Grimus:2004cj}
have been proven to be sufficiently realistic as to accommodate solar and atmospheric splittings but 
either the solar angle was too large or the reactor angle was (almost) vanishing. With the increasing precision in the measurement of oscillation parameters, it turned out that both $\theta_{12}$ and $\theta_{13}$ were substantially different from their LO results; the observation that corrections of ${\cal O}(\lambda)$ ($\lambda$ being the Cabibbo angle) are needed to bring both mixing angles to their experimental values, encouraged to explore the contributions to the neutrino mixing matrix from the charged lepton sector \cite{Petcov:2004rk,Altarelli:2005pj}; in this context, a natural value of $r=\Delta m^2_{\it sol}/\Delta m^2_{\it arm}\sim {\cal O}(\lambda^2)$ was also obtained \cite{Meloni:2011ac}, thus showing that models based on $L_e - L_\mu - L_\tau$ are capable to successfully describe low energy neutrino data. 
An important missing piece of the previous constructions is the possibility to explain the value of the BAU through leptogenesis. In \cite{Grimus:2004cj} it was clearly shown that the baryon-to-photon ratio of the Universe $\eta_B$ is proportional to the neutrino mass $m_1$ and, for vanishing lightest mass $m_3$, $m_1 \sim \sqrt{\Delta m^2_{\it atm}}$, thus producing an hopelessly small $\eta_B$.
Providing a quantitative leptogenesis analysis has become sophisticated in recent years, due to the addition of many ingredients, such as various washout effects \cite{Buchmuller:2004nz,Giudice_2004} or thermal corrections to the particle masses \cite{Giudice_2004,Comelli:1996vm}. Also, the flavor effects can have a significant impact on the final value of the baryon asymmetry, as widely shown in \cite{Abada:2006fw,Nardi:2006fx,Fong:2012buy}.\\
With the present paper we aim to go beyond the existing literature, assessing whether 
see-saw models based on the $L_e - L_\mu - L_\tau$ quantum number can simultaneously account for neutrino masses and mixing and explain the BAU through thermal leptogenesis. 
The paper is structured as follows. In Sect.\ref{model} we describe our model and derive the analytic expressions for the mass ratio $r$ and the mixing angles, showing that appropriate choices of the Lagrangian parameters lead to a satisfactory description of low energy data; in Sect.\ref{lepto} we face the problem of reproducing the value of $\eta_B$, analyzing the resonant and hierarchical scenarios and solving the related Boltzmann equations. Our conclusions are drawn in Sect.\ref{concl}.
\section{The Model}
\label{model}
\begin{table}[t]\centering
\centering
\begin{adjustbox}{width=0.5\textwidth}
\small
\begin{tabular}{|c|c|c|c|c|c|c|c|c|c|c|c|c|c|}
\hline
 & $l_e$& $l_\mu$ & $l_\tau$ & $l_e^c$& $l_\mu^c$&$l_\tau^c$&$F_1$&$F_2$&$H_u$&$H_d$&$N_1$&$N_2$&$N_3$ \\
\hline
\hline
U(1$)_\text{F}$& +1 & -1&-1&$-13$&$7$&3&2&1/2&0&0&-1&1&0\\
\hline
\end{tabular}
\end{adjustbox}
\caption{\it $U(1)_F$ charges for leptons, Higgses and flavon fields.}
\label{charges}
\end{table}
In the following, we summarize the relevant features of our see-saw flavor model based on a broken $U(1)_F$ symmetry.
In the proposed scenario, the left handed lepton doubles
have charge $L_e-L_\mu-L_\tau$ \cite{Altarelli:2005pj} under the $U(1)_F$, while the right-handed $SU(2)$ singlets $l_{e,\mu,\tau}^c$ have the charges reported in Tab.\ref{charges}. The spectrum of the theory also contains three heavy sterile neutrinos $N_{i=1,2,3}$, needed for the generation of the light neutrino masses as well as for the implementation of the leptogenesis process. The flavor symmetry is broken by vacuum expectation values (vev's) of $SU(2)$ singlet scalar fields (flavons) suitably charged under the $U(1)_F$ symmetry. Non-vanishing vevs are determined by the D-term potential \cite{Altarelli:2008bg}:
\begin{eqnarray}
V_D=\frac{1}{2}(M_{\text{FI}}^2-g_F|F_1|^2-g_F|F_2|^2)\,, \end{eqnarray}
where $g_F$ denotes the gauge coupling constant of the $U(1)_F$ symmetry while $M_{\text{FI}}$  is the Fayet-Iliopulos term. Non-zero vevs are obtained by imposing the SUSY minimum $V_D=0$. Without loss of generality, we can assume equal vevs for the two flavons and define $\lambda=\langle F_1 \rangle /M_F=\langle F_2 \rangle /M_F$ the common ratio between the vevs of the flavons and the scale $M_F$ at which the flavour symmetry is broken. 
\subsection{Charged lepton sector}
In the charged lepton sector, many operators of different dimensions enter the Lagrangian; to avoid cumbersome expressions, we quote here the lowest dimensional operators contributing to  each entry of the mass matrix:
\begin{equation}
\begin{aligned}
\mathcal{L}=&a_{11}l_el_e^c\left( \frac{F_1}{M_F}\right)^6H_d+a_{12}l_el_\mu^c\left( \frac{F_1^\dagger}{M_F}\right)^4H_d+a_{13}l_el_\tau^c\left( \frac{F_1^\dagger}{M_F}\right)^2H_d+\\
&a_{21}l_\mu l_e^c\left( \frac{F_1}{M_F}\right)^7H_d+a_{22}l_\mu l_\mu^c\left( \frac{F_1}{M_F}\right)^3H_d+a_{23}l_\mu l_\tau^c\left( \frac{F_1^\dagger}{M_F}\right)H_d+\\
&a_{31}l_\tau l_e^c\left( \frac{F_1}{M_F}\right)^7H_d+a_{32}l_\tau l_\mu^c\left( \frac{F_1}{M_F}\right)^3H_d+a_{33}l_\mu l_\tau^c\left( \frac{F_1^\dagger}{M_F}\right)H_d+\text{h.c.}\,,
\end{aligned}
\end{equation}
where all $a_{ij}$ coefficients are generic ${\cal O}(1)$ free parameters.
After flavor and electroweak symmetry breakings, the previous Lagrangian generates a mass matrix whose elements have the general structure:
\begin{equation}
    (m_l)_{ij}\sim a_{ij}l_i l_i^c {\left(\frac{\langle F_1 \rangle}{M_F}\right)}^{\alpha_{ij}}{\left(\frac{\langle F_2 \rangle}{M_F}\right)}^{\beta_{ij}}\langle H_d\rangle\,,
\end{equation}
where $\alpha_{ij}$ and $\beta_{ij}$ denote the appropriate powers of the flavon fields needed to generate a singlet under $U(1)_F$. 
Factorizing out the $\tau$ mass, the charged lepton mass matrix assumes the following form: 
\begin{equation}
m_l\sim m_\tau\begin{pmatrix}a_{11}\lambda^5&a_{12}\lambda^3&a_{13}\lambda\\
a_{21}\lambda^6&a_{22}\lambda^2\text{e}^{i\phi_{22}}&a_{23}\text{e}^{i\phi_{23}}\\	a_{31}\lambda^6&a_{32}\lambda^2\text{e}^{i\phi_{32}}&1\end{pmatrix}\,.
\end{equation}
For $\lambda <1$, the following mass ratios $m_e:m_\mu:m_\tau=\lambda^5:\lambda^2:1$ is found, which naturally reproduces the observed pattern if $\lambda \sim 0.22$. 
It is not difficult to derive the left-handed rotation $U_l$, which contributes to the total neutrino mixing matrix $U_{PMNS}$ 
\cite{1962PThPh..28..870M,Pontecorvo:1957qd}, diagonalizing the hermitean $m_l m_l^\dagger$ combination:
\begin{equation}
U_l=\begin{pmatrix}-1&\dfrac{a_{13}\lambda}{\sqrt{1+a_{23}^2e^{2i\phi_{23}}}}&\dfrac{a_{13}\lambda}{\sqrt{1+a_{23}^2e^{2i\phi_{23}}}}\\
-\dfrac{a_{13}a_{32}e^{i(\phi_{22}+\phi_{23})}\lambda}{-a_{23}a_{32}e^{i\phi_{22}}+a_{22}e^{i(\phi_{23}+\phi_{32})}}&\dfrac{a_{23}e^{i\phi_{23}}}{\sqrt{1+a_{13}^2e^{2i\phi_{23}}}}&\dfrac{a_{23}e^{i\phi_{23}}}{\sqrt{1+a_{13}^2e^{2i\phi_{23}}}}\\
-\dfrac{a_{13}a_{22}e^{i(\phi_{23}+\phi_{32})}\lambda}{-a_{23}a_{32}e^{i\phi_{22}}+a_{22}e^{i(\phi_{23}+\phi_{32})}}&\dfrac{1}{\sqrt{1+a_{23}^2e^{2i\phi_{23}}}}&\dfrac{1}{\sqrt{1+a_{23}^2e^{2i\phi_{23}}}}\end{pmatrix}+\mathcal{O}(\lambda^2)\,.
\end{equation}
As expected, the diagonal elements of $U_l$ are unsuppressed; also, the (23) and (33) entries are of ${\cal O}(1)$, which indicates that the charged lepton contribution to the atmospheric angle will be large. Notice also that, being the (12) and (13) elements of ${\cal O}(\lambda)$, we expect a similar corrections to the solar and reactor angles.
\subsection{Neutrino sector}
In the neutrino sector, masses are generated through the standard type-I see-saw mechanism;
at the renormalizable level, 
the see-saw Lagrangian reads:
\begin{equation}
\begin{aligned}
\mathcal{L}^{LO}=&\frac{1}{2} \mathcal{M} W \overline{N}^c_1N_2+\frac{1}{2} \mathcal{M} Z \overline{N}^c_3N_3 - a\overline{N}_1H_ul_\mu+\\
&-b\overline{N}_1H_ul_\tau-c\overline{N}_2H_ul_e+\text{h.c.   }\,,
\end{aligned}
\end{equation}
where  $\mathcal{M}$ is an overall Majorana mass scale while $W,Z,a,b,c$ are dimensionless coefficients which will be regarded as free parameters. When $H_u$ acquires a vev $v_u$, Majorana and Dirac mass matrices are generated:
\begin{equation}
\label{renor}
    M_R=\mathcal{M}\left(\begin{array}{ccc}
    0  &  W & 0  \\
    W   &  0 & 0 \\
    0  &  0 & Z
    \end{array}
    \right),\,\,\,m_D=v_u \left(\begin{array}{ccc}
    0   & a & b \\
    c   & 0 & 0 \\
    0  &  0 & 0
    \end{array}
    \right)\,.
\end{equation}
Next-to-leading order (NLO) contributions are given by higher dimensional operators suppressed by the large scale $M_F$; up to two-flavon insertions, we get\footnote{Notice that we use the same overall  scale $\mathcal{M}$ in the Majorana mass terms.}:
\begin{equation}
    \begin{aligned}
    \mathcal{L}^{NLO}=&\frac{1}{2}\mathcal{M} m_{11}\overline{N}_1^cN_1\left( \frac{ F_1}{M_F}\right)+\frac{1}{2} \mathcal{M} m_{13}\overline{N}_1^cN_3\left( \frac{ F_2}{M_F}\right)^2+\frac{1}{2} \mathcal{M} m_{22}\overline{N}_2^cN_2\left( \frac{ F_1^\dagger}{M_F}\right)+\\
    &+\frac{1}{2}\mathcal{M} m_{23}\overline{N}_2^cN_3\left( \frac{ F_2^\dagger}{M_F}\right)^2-d_{11}\overline{N}_1H_ul_e\left( \frac{ F_1^\dagger}{M_F}\right)-d_{22}\overline{N}_2H_ul_\mu\left( \frac{ F_1}{M_F}\right)+\\
    &-d_{23}\overline{N}_2H_ul_\tau\left( \frac{ F_1}{M_F}\right)-d_{31}\overline{N}_3H_ul_e\left( \frac{ F_2^\dagger}{M_F}\right)^2-d_{32}\overline{N}_3H_ul_\mu\left( \frac{ F_2^\dagger}{M_F}\right)^2+\\
    &-d_{33}\overline{N}_3H_ul_\tau\left( \frac{ F_2^\dagger}{M_F}\right)^2+\text{h.c.}\,.
    \end{aligned}
    \label{lnlo}
\end{equation}
Their main effects is to fill the vanishing entries in eq.(\ref{renor}); however, as we have numerically verified, some of the free parameters in eq.(\ref{lnlo}) needed to be slightly adjusted to fit the low energy data. In particular, only a moderate fine-tuning is necessary on $m_{11},\,m_{22},\,d_{11},\,d_{22}$ and $d_{23}$, which we rescale according to:
\begin{eqnarray}\nonumber
\label{yuk}
(m_{11},m_{22},d_{11},d_{22},d_{23})\rightarrow \lambda\; (m_{11},m_{22},d_{11},d_{22},d_{23})\,,\qquad 
(m_{11},m_{22},d_{11},d_{22},d_{23})\sim {\cal{O}}(1)\,.
\end{eqnarray}
With the previous position (and reminding that $\lambda=\langle F_1 \rangle /M_F=\langle F_2 \rangle /M_F$), the following Dirac and Majorana mass matrices are obtained:
\begin{equation}
Y=\frac{m_D}{v_u}\sim \overline{N}l\sim\begin{pmatrix}
                \lambda^2 d_{11}&a e^{i\Sigma}&be^{i\Omega}\\
                c e^{i\Phi}&\lambda^2 d_{22}&\lambda^2 d_{23}e^{i\Theta}\\
                \lambda^2 d_{31}&\lambda^2 d_{32}&\lambda^2 d_{33}
                \end{pmatrix}\,,
\label{matrixDirac}
\end{equation}
and 
\begin{equation}
M_R\sim \overline{N}N\sim \mathcal{M}\begin{pmatrix}\lambda^2 m_{11}&W&\lambda^2 m_{13}\\
					W&\lambda^2 m_{22}&\lambda^2m_{23}\\
					\lambda^2m_{13}&\lambda^2m_{23}&Z\end{pmatrix}\,.
					\label{matrixMajo}
\end{equation}
Notice that the Dirac mass matrix contains un-suppressed entries because of the choice $Q_{N_1}=-Q_{N_2}$ for two of the right-handed neutrinos. The four physical phases $\Sigma, \Omega, \Phi, \Theta$ in $Y$, obtained after a suitable redefintion of the fermion fields, 
are the only source of CP violation of our model and are not fixed by the symmetries of the Lagrangians.
For the sake of simplicity and without any loss of generality, we can assume the parameters $m_{ij}\sim m$ and consider $m$ as a real quantity.
From the type-I seesaw master formula, $m_\nu\simeq -v_u^2Y^\text{T}M_R^{-1}Y$, we get the following matrix for the light SM neutrinos, up to $\mathcal{O}(\lambda^2)$:
\begin{equation}
\label{eq:m_nu}
\begin{aligned}
&m_\nu \simeq
\frac{v_u^2}{\mathcal{M}}\times\\
&\times\begin{pmatrix}\frac{ce^{i\Phi}\left( ce^{i\Phi}m-2d_{11}W\right)\lambda^2}{W^2}&\bigcdot&\bigcdot\\
-\frac{ace^{i(\Sigma+\Phi)}}{W}&\frac{ae^{i\Sigma}\left(ae^{i\Sigma}m-2d_{22}W \right)\lambda^2}{W^2}&\bigcdot\\
-\frac{bce^{i(\Phi+\Omega)}}{W}&\frac{\left( abe^{i(\Sigma+\Omega)}m-ad_{23}e^{i(\Theta+\Sigma)}W-bd_{22}e^{i\Omega}W\right)\lambda^2}{W^2}&\frac{be^{i\Omega}\left(be^{i\Omega}m-2d_{23}e^{i\Theta}W\right)\lambda^2}{W^2}
\end{pmatrix}\,.
\end{aligned}
\end{equation}
This  mass matrix provides, as usual for models based on the  $L_e-L_\mu-L_\tau$ symmetry, an inverted mass spectrum.\\
From now on, we will distinguish two different scenarios, that will be further elaborated when studying the BAU generated in our model, and identified by different assumptions on the parameter $\mathcal{M}$. In order to better understand this distinction, let us assume that all parameters in eq.(\ref{eq:m_nu}) are of $\mathcal{O}(1)$; thus, the light neutrino mass matrix can be recast in the following form:
\begin{equation}
\label{mnu_simpl}
m_\nu=m_0\begin{pmatrix}\lambda^2 x_1&1&x\\
					1&x_2\lambda^2&x_3\lambda^2\\
					x&x_3\lambda^2&x_4\lambda^2\end{pmatrix}\,,
\end{equation}
where $m_0=v_u^2/\mathcal{M}\times {\cal O}(1)$ coefficients and $(x,x_i)$ are suitable combinations of the coefficients present in Dirac and Majorana matrices in eqs.(\ref{matrixDirac}) and (\ref{matrixMajo}). At the leading order in $\lambda$, $m_\nu$ has two degenerate eigenvalues $m_1=-m_2=\sqrt{1+x^2}$ and a vanishing one, $m_3=0$; therefore, we can only construct the atmospheric mass difference $\Delta m_{atm}^2=|m_1|^2-|m_3|^2$, which results in:
\begin{equation}
    x^2=\frac{\Delta m_{atm}^2}{m_0^2}-1\,.
\end{equation}
To maintain $x\sim \mathcal{O}(1)$, we can choose the overall mass scale to $m_0\sim \mathcal{O}(10^{-2})$ eV, which corresponds to the choice $\mathcal{M}\sim 10^{15}$ GeV. 
Notice also that, taking into account the corrections of $\mathcal{O}(\lambda^2)$, the eigenvalue degeneracy is broken and the solar mass difference can be accounted for, which results in:
\begin{equation}
        \Delta m^2_{sol}=\dfrac{ \lambda^2\left[x_1(1+x^2)+x_2+2xx_3+x_4x^2\right]}{\sqrt{1+x^2}}\,.
\end{equation}
Since the masses of the three heavy right-handed neutrinos are simply given by $M_i = \mathcal{M} \widetilde{M}_i$ with:
\begin{equation}
\begin{aligned}
\widetilde{M}_{1}\simeq &\;W+ m\lambda^2+\mathcal{O}(\lambda^3),\\
\widetilde{M}_{2}\simeq &\;W-m\lambda^2+\mathcal{O}(\lambda^3),\\
\widetilde{M}_3\simeq &\;Z + \mathcal{O}(\lambda^3)\,,
\end{aligned}
\end{equation}
and, in particular, the relation  $W \simeq Z$ holds, we dubbed such a situation as the  {\it resonant scenario}. As expected, two mass eigenstates are degenerate, up to corrections of order $\lambda^2$. 
The second possibility arises when  $\mathcal{M}\sim 10^{13}$ GeV and, consequently, $W$ should be around  $10^2$ to maintain $m_0\sim \mathcal{O}(10^{-2})$ eV. With all other parameters again of ${\cal O}(1)$, the light neutrino mass matrix is now as follows:
\begin{equation}
m_\nu\simeq
\frac{v_u^2}{\mathcal{M}W}\begin{pmatrix}-2ce^{i\Phi}d_{11}\lambda^2&-ace^{i(\Sigma+\Phi)}&-bce^{i(\Phi+\Omega)}\\
-ace^{i(\Sigma+\Phi)}&-2ae^{i\Sigma}d_{22}\lambda^2&-\left(ad_{23}e^{i(\Theta+\Sigma)}+bd_{22}e^{i\Omega}\right)\lambda^2\\
-bce^{i(\Phi+\Omega)}&-\left(ad_{23}e^{i(\Theta+\Sigma)}+bd_{22}e^{i\Omega}\right)\lambda^2&-2bd_{23}e^{i(\Theta+\Omega)}\lambda^2
\end{pmatrix}\,.
\label{eq:lownu}
\end{equation}
The right-handed neutrino spectrum will feature two, almost degenerate states with mass $M_1\simeq M_2 \sim 10^{15}$ GeV, and a lighter one with mass $M_3\simeq Z\mathcal{M}\sim 10^{13}$ GeV; we call this scenario as the {\it hierarchical scenario}. Even in this case, the left-handed neutrino mass matrix can be recast in the form (\ref{mnu_simpl}), with the obvious redefinition of  $(x,x_i)$ in terms of the parameters appearing in eq.(\ref{eq:lownu}). Thus, in both resonant and hierarchical scenarios, we get the same structure of the diagonalizing matrix $U_\nu$: 
\begin{equation}
U_\nu=\begin{pmatrix}-\dfrac{1}{\sqrt{2}}&\dfrac{1}{\sqrt{2}}&0\\
\dfrac{1}{\sqrt{2(1+x^2)}}&\dfrac{1}{\sqrt{2(1+x^2)}}&-\dfrac{x}{\sqrt{2(1+x^2)}}\\
\dfrac{x}{\sqrt{2(1+x^2)}}&\dfrac{x}{\sqrt{2(1+x^2)}}&-\dfrac{1}{\sqrt{1+x^2}}\end{pmatrix}+\mathcal{O}(\lambda^2).
\end{equation}
\begin{table}[t]
\centering
\renewcommand{\arraystretch}{1.6}
\begin{tabular}{|c|c|}
\hline
\textbf{Oscillation parameters}              & \textbf{Best fits (IH)}    \\ \hline
$\theta_{12}/^\circ$                         & $33.45^{+0.78}_{-0.75}$    \\ \hline
$\theta_{13}/^\circ$                         & $8.61^{+0.12}_{-0.12}$     \\ \hline
$\theta_{23}/^\circ$                         & $49.3^{+1.0}_{-1.3}$       \\ \hline
$\delta_{cp}/^\circ$                         & $287^{+27}_{-32}$       \\
\hline
$\frac{\Delta m_{21}^2}{10^{-5} \, \, eV^2}$ & $7.42^{+0.21}_{-0.20}$     \\ \hline
$\frac{\Delta m_{31}^2}{10^{-3} \, \, eV^2}$ & $-2.497^{+0.028}_{-0.028}$ \\ \hline
\end{tabular}
\caption{\label{SMparam}\it Best fits and 1$\sigma$ ranges for the oscillation parameters, from \cite{2020}.}
\label{tablenufit}
\end{table}
The final expressions of the neutrino mixing angles (from the relation $U_{PMNS}=U_l^\dagger U_\nu$) and the ratio $r=\Delta m^2_{sol}/\Delta m^2_{atm}$ in terms of the model parameters are quite cumbersome. Thus, we prefer to report here the order of magnitude in the expansion parameter $\lambda$, which are valid for both scenarios considered in this paper: 
\begin{equation}
    \begin{aligned}
    r\sim&\, \mathcal{O}(\lambda^2),\\
    \tan \theta_{12}\sim&\,1+\mathcal{O}(\lambda),\\
    \sin \theta_{13}\sim &\,\mathcal{O}(\lambda),\\
    \tan \theta_{23} \sim &\, 1\,.
    \label{eq:fit}
    \end{aligned}
\end{equation}
Our order of magnitude estimates are in good agreement with their measured values, reported in Tab.\ref{tablenufit}. This conclusion has been further strengthened by a succesful  numerical scan \cite{Marciano:2021oah}  
over the model free parameters, with moduli extracted flat in the intervals $[\lambda,5]$ and all the phases in $ [-\pi,\pi]$. 
Notice that the model provides a CP conserving leptonic phase\footnote{We preferred not to report here its analytic cumbersome expression.} (still compatible with the data at less than 3$\sigma$).
\section{Leptogenesis}
\label{lepto}
As stated above, our study of leptogenesis will be performed within two reference scenarios, identified by different mass patterns for the heavy right-handed neutrinos:
\begin{itemize}
\item {\it{the resonant scenario}}, with $M_1\simeq M_2\simeq M_3$ around $10^{15}$ GeV, and the mass difference comparable to the decay width $\Delta M_{ij}\sim \Gamma_i$\,;
\item {\it{the hierarchical scenario}}, with $M_3\ll M_{1,2}\simeq 10^{15}$ GeV\,.
\end{itemize}
The three Majorana neutrinos decay in the early Universe creating a lepton asymmetry, which is consequently conversed in a baryon asymmetry through non perturbative processes, known as \emph{sphaleron processes} \cite{Giudice_2004,Buchmuller:2004nz}. As we will clearify in the following, a different  Majorana neutrino mass spectrum can lead to different CP-violating parameters, affecting the final amount of baryon asymmetry in the Universe.
\subsubsection*{Resonant Scenario}
We are interested in computing the BAU via thermal leptogenesis.
To facilitate the understanding of the numerical results, we will first provide analytical (approximated) expressions of all relevant quantities entering our computations, which will be validated against a full numerical solution of an appropriate system of Boltzmann's equations. 
Being mass degenerate, we expect that all the three right-handed neutrinos contribute to the leptogenesis process. Following \cite{Davidson:2008bu}, we write the baryon asymmetry as:
\begin{equation}
\eta_B\simeq 7.04\cdot 10^{-3}\sum_{i}\varepsilon_i\eta_i\quad \text{with i=1,2,3}\,,
\label{eqn:etaBResonance}
\end{equation}
with $\varepsilon_i$ being the CP-asymmetries produced in the decay of the i-th neutrino and $\eta_i$ the corresponding efficiency factors. As will be clarified below, the generation of the lepton asymmetry in the resonant scenario occurs in an unflavored regime; the final asymmetry is consequently just the sum of the contributions associated to the individual neutrinos.
The efficiency factors can be written in terms of decay parameters $K_i$ \cite{Fong:2012buy}:
\begin{equation}
    \eta_i=\frac{1}{K_i},\,\,\,\,\,K_i\equiv \frac{\widetilde{m}_i}{m^\ast}\,,
    \label{efficiencyfactors}
\end{equation}
with:
\begin{equation}
\begin{aligned}
\widetilde{m}_i=&\frac{Y^\ast_{i i}Y_{i i} v_u^2}{M_i}\,,\\
m_\ast=&1.1\times 10^{-3}\text{ eV}\,.
\end{aligned}
\label{eq:tilde}
\end{equation}
As it is well known, different values of $K_i$ define different washout regimes, namely \emph{strong} ($K_i \gg 1$), \emph{intermediate} ($K_i \simeq 1$) and \emph{weak} ($K_i \ll 1$). In terms of the parameters of the Dirac and Majorana neutrino mass matrices of eq.(\ref{matrixDirac}), their expressions up to ${\cal O}(\lambda^4)$ are as follows:
\begin{equation}
\begin{aligned}
K_1\simeq& \dfrac{27.7}{W} \bigg\{c^2+\left( d_{11}^2+d_{31}^2\right)\lambda^4+\frac{2}{(W-Z)}\left( \frac{c^2m^2}{(W-Z)}+\sqrt{2}\,c\,d_{31}\,m\cos( \Phi)\right)\lambda^4\bigg\}\,,\\
K_2&\simeq \dfrac{27.7}{W} \big\{a^2 +(d_{22}^2+d_{33}^2)\lambda^4\big\},\quad K_3\simeq\dfrac{27.7}{Z} \big\{b^2 + (d_{23}^2+d_{33}^2)\lambda^4\big\}\,.
\end{aligned}
\end{equation}
Being all the above parameters of 
${\cal O}(1)$, we straightforwardly conclude $K_i \sim {\cal O}(10)$, implying a intermediate/strong washout regime.\\ 
We are now in the position to discuss the CP asymmetry.
Since the heavy neutrinos are close in mass, the CP-asymmetry can be resonantly enhanced \cite{Davidson:2008bu,Pilaftsis:2003gt,Dev_2018}.
To check whether such an enhancement occurs, and hence properly evaluate the CP asymmetry parameters, a good rule of thumb consists in computing
the ratios $\Delta M_{ij}/\Gamma_i$ between the mass splittings and the decay widths of the right-handed neutrinos, verifying that the resonance condition $\Delta M_{ij}\sim \Gamma_i$ is satisfied. In the scenario under scrutiny,  with degenerate masses and not strongly hierarchical Yukawa couplings, we can assume $\Gamma_1\sim \Gamma_2\sim \Gamma_3$, so that:
\begin{equation}
\begin{aligned}
\frac{\Delta M_{12}}{\Gamma _{1}}&\simeq\frac{\Delta M_{12}}{\Gamma _{2}}\simeq \frac{32\,m\pi}{a^2\,W}\lambda^2+{\cal O}(\lambda^4)\;,\\
\frac{\Delta M_{23}}{\Gamma _{2}}&\simeq\frac{\Delta M_{23}}{\Gamma _{3}}\simeq \frac{16\,(Z-W)\,\pi }{b^2+\,Z}+{\cal O}(\lambda^2)\;,\\
\frac{\Delta M_{13}}{\Gamma _{1}}&\simeq\frac{\Delta M_{13}}{\Gamma _{3}}\simeq  \frac{16\,(Z-W)\,\pi }{b^2+\,W}+{\cal O}(\lambda^2)\;,
\end{aligned}
\end{equation}
where we have used $\Gamma_i=M_i\left( Y^\dagger Y\right)_{ii}/(16 \pi )$.
As evident, for $W\hspace{-0.14cm}-\hspace{-0.14cm}Z\simeq O(0.1)$, $\Delta M_{i3}\simeq \Gamma_{i}$. Consequently, leptogenesis occurs in the resonant regime. In such a case, the self energy contribution dominates the CP violation parameters. Furthermore, as shown in \cite{DeSimone:2007edo}, the asymmetry parameters are time dependent. Following \cite{DeSimone:2007edo,DeSimone:2007gkc}, we rewrite the latter as:   
\begin{equation}
\label{eq:eps_res1}
\varepsilon_i(z)\simeq \sum_{j\not =i} \frac{\text{Im}\left[ \left( Y^\dagger Y\right)^2_{ij}\right]}{(Y^\dagger Y)_{ii}(Y^\dagger Y)_{jj}}\frac{\Delta M_{ij}/\Gamma_{j}}{1+\left(\Delta M_{ij}/\Gamma_{j}\right)^2}\left[ f_{ij}^\text{mix}(z)+f_{ij}^\text{osc}(z)\right]\,,
\end{equation}
where $z=M/T$, with $M$ the sterile neutrino mass.
The coefficient in front of the squared parenthesis is the (constant) usual CP-asymmetry and it is resonantly enhanced for $\Delta M\sim \Gamma$, while the second one is the sum of two $z$ (and hence time) dependent functions:
\begin{equation}
\label{eq:eps_res2}
\begin{aligned}
f^{\text{mix}}_{ij}(z)=& 2 \sin^2 \left(\Delta M_{ij} t\right)=+2\sin^2\left[\frac{K_i z^2 \Delta M_{ij}}{4\Gamma_{i}} \right],\\
f^{\text{osc}}_{ij}(z)=&- \frac{\Gamma_j}{\Delta M_{ij}} \sin \left(\Delta M_{ij} t\right)=-\frac{\Gamma_j}{\Delta M_{ij}}\sin\left[ \frac{K_i z^2\Delta M_{ij}}{2\Gamma_i}\right],
\end{aligned}
\end{equation}
In the case $\Delta M_{ij}t \equiv \frac{K_i z^2 \Delta M_{ij}}{4\Gamma_i} \gg 1$,  eq.(\ref{eq:eps_res2}) is a strongly oscillating function. Making the average over a generic time interval (or, equivalently, in $z$) $t\in [0,\tau]$, we have that:
\begin{equation}
    \left \langle 2\sin^2 \Delta M_{ij}t-\frac{\Gamma_{ij}}{\Delta M_{ij}}\sin \Delta M_{ij}t\right \rangle=1-\frac{\sin 2 \Delta M_{ij}\tau}{2 \Delta M_{ij}\tau}-\frac{\Gamma_i}{\Delta M_{ij}}\frac{1-\cos \Delta M_{ij}\tau}{\Delta M_{ij}\tau}
\end{equation}
 As evident, the average tends to 1 if $\Delta M_{ij}\tau \gg 1$. In such a limit we have that the CP asymmetry averages to the following constant value:
\begin{equation}
    \langle \epsilon(z) \rangle=\sum_{j\not =i} \frac{\text{Im}\left[ \left( Y^\dagger Y\right)^2_{ij}\right]}{(Y^\dagger Y)_{ii}(Y^\dagger Y)_{jj}}\frac{\Delta M_{ij}/\Gamma_{j}}{1+\left(\Delta M_{ij}/\Gamma_{j}\right)^2}\,.
\end{equation}
In the case of a large decay parameter, the regime $\Delta M_{ij} t \gg 1$ occurs for small values of $z$. In good approximation the whole leptogenesis process can be described by replacing the time dependent CP-asymmetry with its average. Being, in our case, $K_i \simeq {\cal{O}}(10)$, we can adopt this approach and, consequently, propose an analytic estimate of the baryon asymmetry neglecting the time dependence of the $\varepsilon_i$'s. The $\eta_B$  will be nevertheless compared against a complete numerical treatment.
The time-independent piece of the CP-asymmetry parameters in eq.(\ref{eq:eps_res1}) can be expressed in power series of the small parameter $\lambda$ as follows:
\begin{equation}
\begin{aligned}
\label{eq:eps_constant}
\varepsilon_1=&+\frac{16\,\pi\,(W-Z)\,Z}{c^2\left( 256\pi(W-Z)^2+b^4Z^2\right)}\big[c^2d_{23}^2\sin\left[ 2\left( \Theta-\Phi\right)\right]+\\
&+bd_{11}\left(bd_{11}\sin\left[ 2\Omega\right]+2cd_{23}\sin \left[ \Theta -\Phi +\Omega\right] \right) \big]\lambda^4+\mathcal{O}(\lambda^5)\\
\varepsilon_2=&-\frac{16\left(b^2\pi(W-Z)Z\sin\left[ 2(\Sigma-\Omega)\right] \right)}{256\pi^2(W-Z)^2+b^4Z^2}+\\
&+\frac{16b^2m\pi Z\left(256\pi^2(W-Z)^2-b^4Z^2 \right)\sin\left[ 2\left(\Sigma-\Omega \right)\right]\lambda^2}{\left( 256\pi^2(W-Z)^2+b^4Z^2\right)^2}+\mathcal{O}(\lambda^4)\\
\varepsilon_3=&+\frac{16\left(b^2\pi(W-Z)Z\sin\left[ 2(\Sigma-\Omega)\right] \right)}{256\pi^2(W-Z)^2+b^4Z^2}+\\
&-\frac{16b^2m\pi Z\left(256\pi^2(W-Z)^2-b^4Z^2 \right)\sin\left[ 2\left(\Sigma-\Omega \right)\right]\lambda^2}{\left( 256\pi^2(W-Z)^2+b^4Z^2\right)^2}+\mathcal{O}(\lambda^4)\,.
\end{aligned}
\end{equation}
\begin{table}\centering
\centering
\begin{adjustbox}{width=0.42\textwidth, height= 1.25cm}
\small
\begin{tabular}{|c|c|c|c|c|c|c|c|}
\hline
 $a$& $b$& $c$ & $d_{11}$ & $d_{22}$& $d_{23}$&$d_{31}$&$d_{32}$\\
\hline
1.28&1.49 & 1.46&1.07&1.13&1.45&1.70&1.01\\
\hline
\hline
 $d_{33}$& $m$ & $\Sigma$&$\Omega$& $\Theta$&$\Phi$&$Z$&$W$\\
\hline
1.50& 1.50 & 0.007&0.007 &-0.003&-0.004&1.38&1.27\\
\hline
\end{tabular}
\end{adjustbox}
\caption{\it Set of parameter assignments which lead to the final baryon asymmetry value shown in fig. (\ref{fig:B-LResonance}). The parameters are selected close to 1 as to avoid particular effects of enhancement or suppression in the $\varepsilon_i$.}
\label{primoResonance}
\end{table}
By inspecting the analytical expressions above we see that, for $W-Z\simeq \mathcal{O}(0.1)$, the parameters $\varepsilon_{2,3}$ tend approximately to $\mathcal{O}(0.1)$, while $\varepsilon_1$ is of order $10^{-4}$. Therefore, the main contribution to the final baryon asymmetry of the Universe is carried by $\varepsilon_{2,3}$, leading to an $\eta_B$ value which exceeds the experimentally favoured one $\eta_B \simeq 6.1 \cdot 10^{-10}$ by several orders of magnitude.
It is nevertheless possible to suppress the values of the $\varepsilon_i$ parameters by an ad-hoc assignations of the phases to trigger a destructive interference among them. First of all, by taking $\Sigma-\Omega \rightarrow 0$, the leading order contributions to $\varepsilon_{2,3}$ go to zero and, as for $\varepsilon_1$, all $\varepsilon$'s are of $\mathcal{O}$$(\lambda^4)$:
\begin{equation}
    \begin{aligned}
     \varepsilon_2=&\frac{32bd_{22}d_{23}\pi(W-Z)Z\sin \left[ \Theta\right]\lambda^4}{a\left( 256 \pi^2(W-Z)^2+b^4Z^2\right)}+\mathcal{O}(\lambda^5)\\[4pt]
    \varepsilon_{3}=&\frac{16\pi W(Z-W)}{b^2}\bigg[\frac{2abd_{22}d_{23}\sin \left[ \Theta\right]}{a^4W+256\pi^2(W-Z)^2}+\\
    +&\frac{b^2d_{11}\sin\left[ 2\Sigma\right]+cd_{23}\left( cd_{23}\sin\left[ 2(\Theta-\Phi)\right]+2bd_{11}\sin\left[ \Theta+\Sigma-\Phi\right]\right)}{c^4W^2+256\pi^2(W-Z)^2}\bigg]\lambda^4+\mathcal{O}(\lambda^5)\,.
    \end{aligned}
\end{equation}
 Upon numerical check, we have found that the $\lambda^4$ suppression was not enough to guarantee viable values of the $\varepsilon_{i}$. Consequently, we need to further assume individually suppressed $\Theta$, $\Phi$ and $\Sigma$ phases. For the benchmark values reported in Tab.\ref{primoResonance}, 
the efficiency factors turn to be:
\begin{equation}
    \begin{aligned}
    \eta_1\simeq 1.0 \cdot 10^{-2}, \quad \eta_2\simeq 2.2 \cdot 10^{-2},\quad \eta_3\simeq 1.6 \cdot 10^{-2}\,,
    \end{aligned}
\end{equation}
while the CP-asymmetry parameters are:
\begin{equation}
    \begin{aligned}
    \label{eq:corr_numbers}
    \varepsilon_1\simeq 1.7 \cdot 10^{-5}, \quad \varepsilon_2\simeq 3.2 \cdot 10^{-6},\quad \varepsilon_3\simeq -1.1 \cdot 10^{-5}\,.
    \end{aligned}
\end{equation}
Using the approximated formula in eq.(\ref{eqn:etaBResonance}), we can estimate the final baryon asymmetry, obtaining $\eta_B\approx 5.20 \cdot 10^{-10}$, in good agreement with the observations.
We conclude our study of the resonant scenario with a numerical validation of the analytical results presented above. We have hence solved the following set of coupled Boltzmann's equations \cite{Buchmuller:2004nz}: 
\begin{equation}
\begin{aligned}
\frac{dN_i}{dz}=&-\left( D_i+S_i\right)\left(N_i-N_i^\text{eq}\right)\quad\;i=1,2,3\\
\frac{dN_{B-L}}{dz}=&\sum_{i}^3\varepsilon_iD_i\left( N_i-N_i^\text{eq}\right)-W_iN_{B-L}\,,
\end{aligned}
\end{equation}
where $N_i$ stands for number density of the RH sterile neutrinos, while $N_{B-L}$ is the amount of $B-L$ asymmetry, both normalized by comoving volume \cite{Buchmuller:2004nz}. $\varepsilon_i=\varepsilon_i(z)$ are the full time dependent asymmetry parameters as given in eqs.(\ref{eq:eps_res1}-\ref{eq:eps_res2}). $D_i$ and $S_i$ indicate, respectively, inverse decay and scattering contributions to the production of the right-handed neutrinos while the $W_i$ represent the total rate of Wash-out processes including both inverse decay and $\Delta L \neq 0$ scattering contributions (see Appendix \ref{appendixA} for further details).
From the benchmark values of Tab.\ref{primoResonance}, the following entries of the Yukawa matrix are obtained:
\begin{equation}
\hat{\text{Y}}=\begin{pmatrix}+0.969-0.004i&-0.343-0.002i&-0.405+0.002i\\
											-1.102+0.004i&-0.408-0.002i&-0.491+0.002i\\
											+0.02851+0.00005i&+1.168-0.008i&+1.355-0.009i\end{pmatrix}\,,
\end{equation}
where the symbol \emph{hat} refers to the Yukawa  evaluated in the physical mass basis of the Majorana neutrinos, \emph{i.e.} $\hat{Y}=UY$, where $U$ is the unitary matrix such that $U^T M_R^{-1}U=\left(M_R^{\text{diag}}\right)^{-1}$.
\begin{figure}
\begin{center} 
  \includegraphics[width=11cm, height=6.5cm]{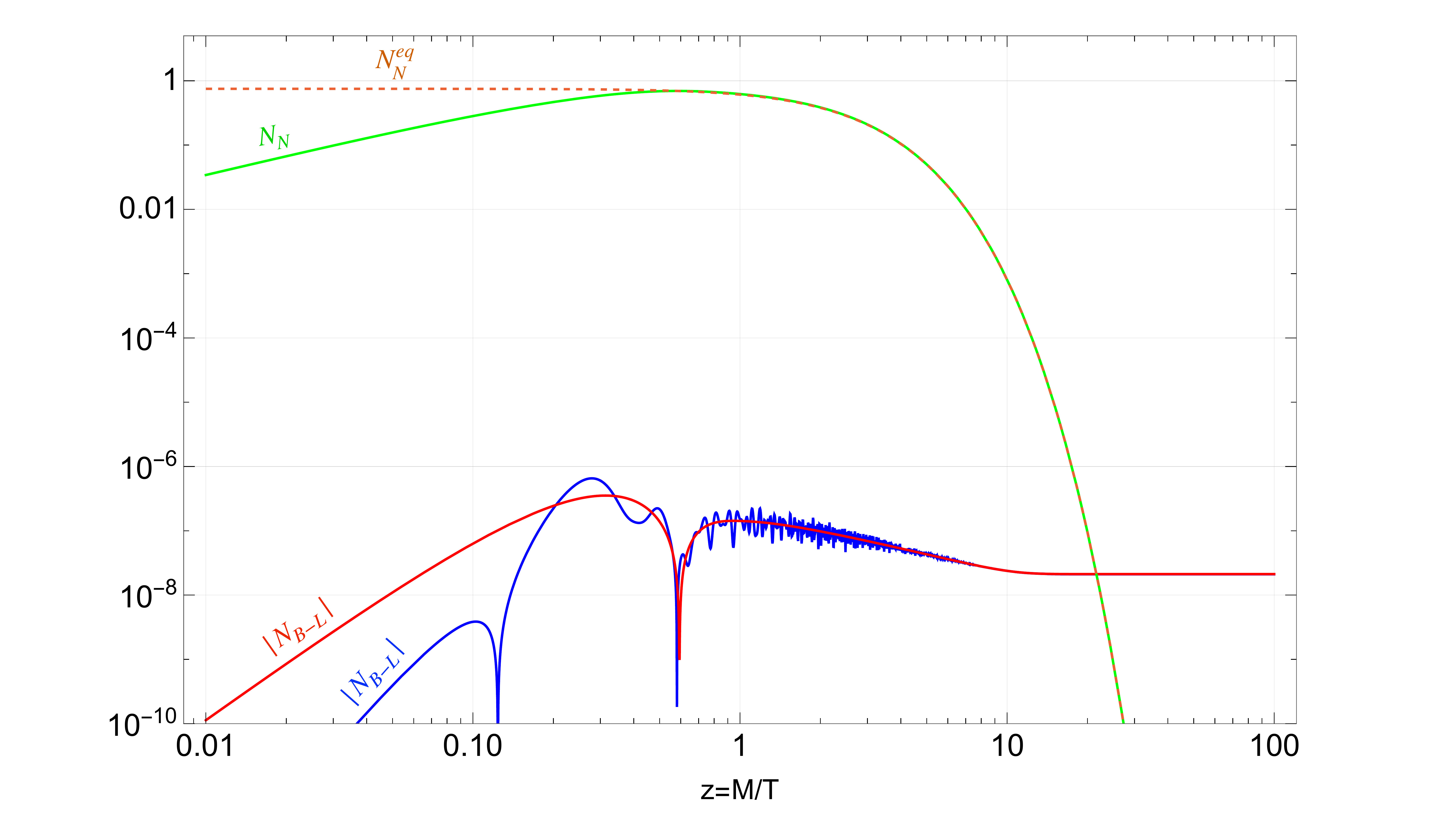}
  \caption{\it $B-L$ asymmetry and neutrino abundance evolution during the expansion of the Universe in case of zero initial neutrino abundance. The blue line refers to the full solution of the Boltzmann's equations, obtained retaining the time dependence of the CP-violation parameter. The red line refers to the solution of the analogous system but adopting a time-constant value of the asymmetry parameters. Finally, the green line represents the abundance of the right-handed neutrinos, as given by the solution of the system. For reference, the latter is compared  with the function $N_{N}^{\rm eq}$ (dashed line) rwhich represents a thermal equilibrium abundance for right-handed neutrinos.}
  \label{fig:B-LResonance}
\end{center} 
\end{figure}
Plugging the latter values in the interaction rates appearing in the Boltzmann's equations, we have solved the system  assuming null initial abundance for the right-handed neutrinos in the primordial plasma. The B-L yield $N_{B-L}$ as a function of $z$ is shown with a blue line in fig.\ref{fig:B-LResonance},  while the abundance of the right-handed neutrinos is displayed with a green line. For reference, we have reported the corresponding equilibrium function as a dashed orange line. To better pinpoint the impact on our result of the time dependency of the CP-asymmetry, we have shown, as red line in fig.\ref{fig:B-LResonance}, solution of the same system of equation but retaining a constant CP violating parameter as given by eq.(\ref{eq:corr_numbers}). As evident, there is a nice agreement between the curves, justifying the assumption of neglecting the time dependence of the asymmetry parameter in our analytical treatment.
Some more comments are in order. Starting from a negligible abundance, the yield of the right-handed neutrino is driven by inverse decays toward the equilibrium value which is reached for $z_{\rm eq}<1$. For $z>z_\text{eq}$, the decays dominate and the neutrino abundance decreases, until it becomes almost zero around $z\simeq 10$. This means that the leptogenesis processes is completed at temperatures above $10^{13}\,\mbox{GeV}$. This justifies our assumption of neglecting flavor effects since the latter are relevant only if leptogenesis occurs at temperatures below $10^{12}\,\mbox{GeV}$. The shape of the $N_{B-L}$ also clearly evidences the time dependency of the $\varepsilon_i$ parameters.
The solution of the system $N_{B-L}(\infty)$ can be related to $\eta_B$ through the relation $\eta_B=(a_{\text{sph}}/f)N_{\text{B-L}}(\infty)$. 
Here $a_{\text{sph}}=28/79$ \cite{Buchmuller:2004nz} is the fraction of $B-L$ asymmetry converted into a baryon asymmetry by the sphaleron processes while $f=N_\gamma^{\text{rec}}/N_\gamma^\ast=2387/86$ is the dilution factor calculated assuming standard photon production from the onset of leptogenesis till recombination. The values of the latter parameters, as obtained from the numerical solution of the system are:
\begin{equation}
\begin{aligned}
N_{B-L}(\infty)=&2.11\times 10^{-8},\\
\eta_{B}=&3.01\times 10^{-10}.
\end{aligned}
\end{equation}
We conclude that, after the phase suppression discussed above, it is possible to provide a viable leptogenesis in the resonant scenario.
\subsubsection*{Hierarchical scenario}
This alternative scenario is obtained by lowering the mass scale $\mathcal{M}$ down to a value of the order of $10^{13}\,\mbox{GeV}$, which brings to a hierarchical mass spectrum for the sterile neutrinos, with two heavy, almost degenerate, states with $M_1\simeq M_2\sim 10^{15}$ GeV, and a lighter one with $M_3\simeq 10^{13}$ GeV.
In this setup, the baryon asymmetry of the Universe can be generated through the conventional thermal leptogenesis only via the out-of-equilibrium decay of the lightest heavy neutrino. Similarly to the previous case, no flavor effects need to be accounted for. 
As discussed in \cite{Davidson:2008bu}, the latter becomes relevant when the rate of processes mediated by $\tau$, $\mu$ and $e$ exceeds the Hubble expansion rate. This occurs when the temperature of the Universe drops below, respectively:
\begin{equation}
\begin{aligned}
 T_\tau\simeq 5\cdot10^{11}\text{ GeV },\quad T_\mu\simeq 2\cdot10^{9}\text{ GeV },\quad T_e\simeq 4\cdot10^{4}\text{ GeV }\,.
\end{aligned}
\end{equation}
In the scenario under consideration we have $M_3\gg T_\tau \gg T_{\mu,e}$. Thus, we can work in the so-called \emph{unflavored regime}, in which the lightest neutrino decays via flavor blind processes. 
The baryon asymmetry can then be parametrized as: 
\begin{equation}
\eta_B\simeq 7.04\cdot 10^{-3}\,\varepsilon_3\, \eta_3\,,
\end{equation}
with $\eta_i$ as in eq.(\ref{efficiencyfactors}) and the unflavored CP-asymmetry as \cite{Blanchet:2012bk,Flanz:1994yx,Covi:1996wh,Buchmuller:1997yu}:
\begin{equation}
\varepsilon_{i}=\frac{1}{8\pi\left( Y^\dagger Y\right)_{ii}}\sum_{j\not=i} \text{Im}\left[\left( Y^\dagger Y\right)_{ij}^2\right]f\left[ \frac{M_j^2}{M_i^2}\right],
\end{equation}
where $x_j=M_j^2/M_i^2$ and the loop function is:
\begin{equation}
f[x]=\sqrt{x}\left[ 1-(1+x)\log\left( 1+\frac{1}{x}\right)+\frac{1}{1-x}\right]\,,
\end{equation}
that can be approximated to $f[x]\simeq -3/(2\sqrt{x})$ in the limit $x\gg 1$.
Expressing the elements of the Yukawa matrices in terms of the model parameters, we have:
\begin{equation}
\begin{aligned}
\varepsilon_{3}=\;f\left[ \frac{M_2^2}{M_3^2}\right]\frac{1}{8b^4\pi}\cdot \bigg \{ &\sin\left[ 2(\Sigma-\Omega)\right]\; \left(a^2b^2-2a^2(d_{23}^2+d_{33}^2)\right)+\\
&-2a\, b\, d_{22}\,d_{23}\sin\left[ \Theta-\Sigma+\Omega\right]+\\
&+2\,a\,b\,d_{32}\,d_{33}\sin\left[\Sigma-\Omega \right] +\mathcal{O}(\lambda^4)\bigg \}.
\end{aligned}
\end{equation}
Using the values in Tab.\ref{estimation_unflavored}, and noticing that $f\left[ M_2^2/M_3^2\right]\simeq 10^{-2}$, we obtain $\varepsilon_3\simeq 3.24\cdot 10^{-4}$.
\begin{table}[t]\centering
\centering
\begin{adjustbox}{width=0.42\textwidth, height= 1.25cm}
\small
\begin{tabular}{|c|c|c|c|c|c|c|c|}
\hline
 $a$& $b$& $c$ & $d_{11}$ & $d_{22}$& $d_{23}$&$d_{31}$&$d_{32}$\\
\hline
1.18&1.07 & 1.65&1.99&1.95&1.88&1.82&1.71\\
\hline
\hline
 $d_{33}$& $m$ & $\Sigma$ & $\Omega$& $\Theta$&$\Phi$&$Z$&$W$\\
\hline
1.83& 1.50 & 0.39& 1.02 & 4.10&5.11&1.41&159.97\\
\hline
\end{tabular}
\end{adjustbox}
\caption{\it Set of parameter assignments which lead to a final baryon asymmetry close to the cosmological observations. See text for further details.}
\label{estimation_unflavored}
\end{table}
The corresponding efficiency factor $\eta_{3}$ can be simply approximated to:
\begin{equation}
    \eta_3\simeq 3.6\cdot 10^{-4}\left(\frac{Z}{b^2} + O(\lambda^4)\right)\simeq 4.3\cdot 10^{-4}\,,
\end{equation}
for the assignations in Tab.\ref{estimation_unflavored}. The efficiency factor can be converted to the baryon asymmetry parameter $\eta_B$:
\begin{equation}
    \eta_B\simeq 7.04\times10^{-3}\left[ \varepsilon_{3}\eta_{3}\right]\approx 9.5 \times 10^{-10}\,,
\end{equation}
very close to the experimentally favoured value.
Also in this case, we have verified the goodness of our analytical approximations by solving a suitable set of Boltzmann equations, which take the form:
\begin{equation}
\begin{aligned}
\frac{dN_3}{dz}=&-\left(D_3-S_3\right)\left( N_3-N_3^\text{eq}\right)\,,\\
\frac{dN_{B-L}}{dz}=&\;\varepsilon_{3}D_3\left( N_3-N_3^\text{eq}\right)-W_3N_{B-L}\,.
\end{aligned}
\end{equation}
From the parameter choices in Tab.\ref{estimation_unflavored}, we get the following Yukawa matrix :
\begin{equation}
\hat{\text{Y}}=\begin{pmatrix}0.39-1.07i&-0.71-0.32i&-0.44-0.70i\\
											-0.52+1.07i&-0.84-0.32i&-0.36-0.59i\\
											+0.08825-0.00004i&+0.0821-0.0003i &+0.0881-0.0007i\end{pmatrix}\,.
\end{equation}
Notice that the imaginary parts of the third row is suppressed by a factor proportional to $Z/W$. \\
This choice corresponds to the CP-violation parameter $\varepsilon_{3}=4.3\cdot 10^{-4}$.
\begin{figure}[tb]
\centering
\includegraphics[width=11cm, height=6cm]{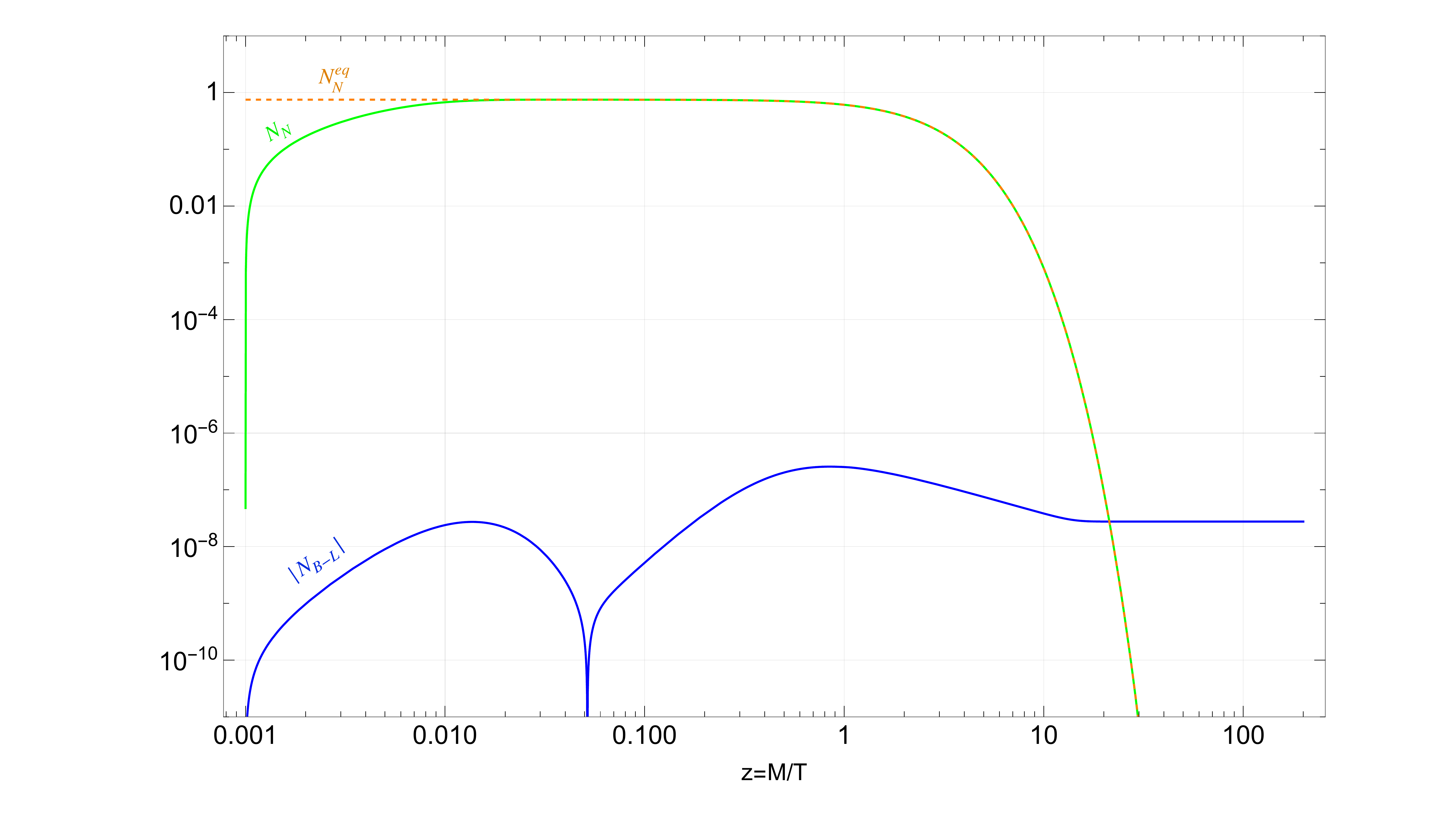}
\caption{\it Evolution of the $B-L$ symmetry in the {\it hierarchical scenario.} The color code is the same as in fig.\ref{fig:B-LResonance}.}
\label{fig:unFlavored}
\end{figure}
For the benchmark values under consideration, the numerical solution of the Boltzmann's equations is shown in fig.\ref{fig:unFlavored} and gives:
\begin{equation}
\begin{aligned}
N_{B-L}(\infty)=&\;2.75\cdot10^{-8}\,,\\
\eta_B=&\;3.96\cdot10^{-10}\,,
\end{aligned}
\end{equation}
confirming what we expected from the simplified analytical analysis.
It is interesting to notice that, contrary to the resonant regime, in the hierarchical scenario it is possible to find a parameter assignation leading to viable leptogenesis without imposing a fine tuning on the CP violating phases.
\section{Conclusions}
\label{concl}
In this paper we have provided a proof of existence about the possibility of contemporary achieving viable masses and mixing patterns for the SM neutrinos and a value of the BAU, via leptogenesis, compatible with the experimental determination, in models based on the abelian flavor symmetry $L_e - L_\mu-L_\tau$. Given the large parameter space of the model, we have identified two reference scenarios. The first one, dubbed the {\it resonant scenario}, provides a viable light neutrino mass spectrum and assures the existence of three degenerate right-handed neutrinos at a mass scale of $10^{15}\,\mbox{GeV}$. In this scenario, the generation of the lepton asymmetry is resonantly enhanced so that a baryon asymmetry exceeding the experimentally favored value is generically predicted. This problem can be overcome by invoking an ad-hoc suppression of the CP-violating phases in the Yukawa matrix. In the second scenario, that we called the {\it hierarchical scenario}, one of the right-handed masses is lowered down to $10^{13}\,\mbox{GeV}$  without destroying the good agreement with the lepton masses and mixing. The BAU is generated via the conventional thermal leptogenesis. We have verified that it is possible to find parameter assignations leading to the correct value of the BAU without invoking specific assignations for the CP violating phases. 
\newpage
\appendix
\section{\\Full Boltzmann Equations}
\label{appendixA}
\hbox{}
\vspace{+0.1 cm}
In this Appendix we discuss in greater detail the Boltzmann's equations solved in the main text.
For the sake of simplicity, we will write them as a single equation for the right-handed neutrino specie $N_1$. In the case of the resonant regime, we just need to consider multiple copies of the same equation.
The relevant processes in the thermal plasma are:
\begin{itemize}
    \item $N_1$ decays (D) and inverse decays (ID) into leptons and Higgs bosons $N_1\rightarrow \phi l$, and into anti-leptons and anti-Higgs bosons $N_1\rightarrow \bar{\phi} \bar{l}$;
    \item $\Delta L=2$ scattering processes mediated by the heavy Majorana neutrinos, $l\phi\leftrightarrow \bar{l}\bar{\phi}$ (N) and $ll\leftrightarrow \bar{\phi}\bar{\phi}$, $\bar{l}\bar{l}\leftrightarrow \phi \phi$ (Nt);
    \item $\Delta L=1$ scattering, with an intermediate Higgs boson field $\phi$, involving the top quark. The s-channel $N_1l\leftrightarrow\bar{t}q$, $N_1\bar{l}\leftrightarrow t\bar{q}$ ($s$) and the t-channel $N_1t\leftrightarrow\bar{l}q$, $N_1\bar{t}\leftrightarrow l\bar{q}$ ($t$);
\end{itemize}
In brackets we have indicated a short-hand notation for such processes, to be used later on. 
Considering $z=M/T$, where $M$ is the mass of the decaying neutrino, the Boltzmann's equations read:
\begin{equation}
    \begin{aligned}
    \frac{dN_1(z)}{dz}=&-\big( D(z)+S(z)\big)\big( N_1(z)-N_1^{eq}(z)\big)\,,\\
    \frac{dN_{B-L}(z)}{dz}=&\varepsilon(z) D(z)\big( N_1(z)-N_1^{eq}(z)\big)-W(z) N_{B-L}(z)\,,
    \end{aligned}
    \end{equation}
    where $\varepsilon$ is the usual CP-violation parameter. Instead of the number density $n_X$ of the particle species, it is useful to consider their number $N_X$ in some portion of the comoving volume, in such a way to automatically take into account the expansion of the Universe. In the literature, the comoving volume $R^3_\ast(t)$ is usually chosen such that it contains one photon at the time $t_\ast$, before the onset of the leptogenesis \cite{Buchmuller:2002rq}:
    \begin{equation}
        N_X(t)=n_X(t)R^3_\ast(t)\, ,
    \end{equation}
    with 
    \begin{equation}
        R_\ast(t_\ast)=\left(n_\gamma^{eq}(t_\ast) \right)^{-1/3}\, ,
    \end{equation}
    so that $N_\gamma(t_\ast)=1$.
   Differently, one could even choose to normalize the number density to the entropy density $s$ considering then $Y_X=n_x/s$, as widely done in literature, \emph{e.g.}, \cite{Buchmuller:2004nz}; however, if the entropy is conserved during the Universe evolution, the two normalizations are related by a constant. 
  Introducing the thermally averaged dilation factor $\braket{1/\gamma}$ as the ratio of the modified Bessel functions of the second type:
   \begin{equation}
       \left \langle\frac{1}{\gamma}\right\rangle=\frac{\mathcal{K}_1}{\mathcal{K}_2}\,,
   \end{equation}
we can write the decay term $D(z)$ as \cite{Kolb:1990vq}:
\begin{equation}
    D(z)=K z \left\langle\frac{1}{\gamma}\right\rangle.
\end{equation}
$K$ is the decay parameter, which is introduced in the context of the GUT baryogenesis \cite{Kolb:1990vq}, to control whether the decays of the sterile neutrinos are in equilibrium or not. This parameter depends on the \emph{effective neutrino mass} $\widetilde{m}_1$ \cite{Plumacher:1996kc}:
\begin{equation}
    \widetilde{m}_1=\frac{\left( Y^\dagger Y\right)_{11}v_u^2}{M_1}\,,
\end{equation}
where $Y$ is the Dirac neutrino Yukawa matrix of eq.(\ref{yuk}), $v_u$ is the vacuum expectation value of the $H_u$ doublet field and $M_1$ is the mass of the decaying neutrino.
\\
This effective mass has to be compared with the \emph{equilibrium neutrino mass} \cite{Buchmuller:2004nz}:
\begin{equation}
    m^\ast\simeq 1.08\cdot 10^{-3}\text{ eV}.
\end{equation}
The decay parameter turns out to be:
\begin{equation}
K=\frac{\widetilde{m}_1}{m^\ast}\;.
\end{equation}
From the parameter $D(z)$ we can obtain the inverse decay parameter $W_{ID}(z)$, which contributes to the washout of the lepton asymmetry. Indeed, the inverse decay parameter can be written as \cite{Buchmuller:2004nz}:
\begin{equation}
    W_{ID}(z)=\frac{1}{2}D(z)\frac{N_1^{eq}(z)}{N_l^{eq}}
\end{equation}
with
\begin{equation}
    N_1^{eq}(z)=\frac{3}{8}z^2\mathcal{K}_2(z)\,,\quad N_l^{eq}=\frac{3}{4}\,.
\end{equation}
Therefore, the contribution of the inverse decays to the final washout is:
\begin{equation}
    W_{ID}(z)=\frac{1}{4}K z^3 \mathcal{K}_1(z)\,.
\end{equation}
We can now move to the $\Delta L=1$ and $\Delta L=2$ scattering processes. The latter contributes to the washout of the lepton asymmetry, while the former counts towards both the production of the right-handed sterile neutrinos and the final washout. 
\\
In general, the scattering terms $S_x(z)$, where the subscript $x$ indicates the different processes to be considered, are:
\begin{equation}
    S_x(z)=\frac{\Gamma_{x}(z)}{Hz}\,,
\end{equation}
with $H$ being the evolution Hubble constant. 
$\Gamma_x$ are the scattering rates, defined as \cite{2009}:
\begin{equation}
    \Gamma_x(z)=\frac{M_1}{24 \zeta(3)g_N\pi^2}\frac{\mathcal{I}_x}{\mathcal{K}_2(z)z^3}\,,
\end{equation}
where $g_N=2$ is the number of degrees of freedom of the right-handed neutrinos. The quantity $\mathcal{I}_x$ is the integral: 
\begin{equation}
    \mathcal{I}_x=\int_{z^2}^\infty d\Psi \hat{\sigma}_x(\Psi)\sqrt{\Psi}\mathcal{K}_1(\sqrt{\Psi})\,,
\end{equation}
of the reduced cross section $\hat{\sigma}_x$ given in \cite{Buchmuller:2002rq}. In particular, for the scattering processes mediated by the three Majorana neutrinos, \emph{i.e.} the $\Delta L=2$ scatterings, the reduced cross section reads \cite{1998}:
\begin{equation}
    \hat{\sigma}_{N,Nt}(x)=\frac{1}{2\pi}\left[\sum_i \left( Y^\dagger Y\right)^2_{ii}f_{ii}^{N,Nt}(x)+\sum_{i<j}\mathcal{R}e\left( Y^\dagger Y\right)^2_{ij}f_{ij}^{N,Nt}(x) \right]\,,
\end{equation}
with 
\begin{equation}
    f_{ii}^N(x)=1+\frac{a_i}{D_i(x)}+\frac{xa_i}{2D_i^2(x)}-\frac{a_i}{x}\left[1+\frac{x+a_i}{D_i}\right]\ln\left( 1+\frac{x}{a_i}\right),
\end{equation}
\begin{equation}
    \begin{aligned}
    f_{ij}^N(x)=\sqrt{a_ia_j}\bigg[&\frac{1}{D_i(x)}+\frac{1}{D_j(x)}+\frac{x}{D_i(x)D_j(x)}+\\
    &+\left( 1+\frac{a_i}{x}\right)\left( \frac{2}{a_j-a_i}-\frac{1}{D_j(x)}\right)\ln\left( 1+\frac{x}{a_i}\right)+\\
    &+\left( 1+\frac{a_j}{x}\right)\left( \frac{2}{a_i-a_j}-\frac{1}{D_i(x)}\right)\ln\left( 1+\frac{x}{a_j}\right)
    \bigg],
    \end{aligned}
\end{equation}
\begin{equation}
    f_{ii}^{Nt}(x)=\frac{x}{x+a_i}+\frac{a_i}{x+2a_i}\ln\left( 1+\frac{x}{a_i}\right),
\end{equation}
\begin{equation}
    \begin{aligned}
    f_{ij}^{Nt}(x)=\frac{\sqrt{a_ia_j}}{(a_i-a_j)(x+a_i+a_j)}\bigg[(2x+3a_i+a_j)\ln\left(1+\frac{x}{a_j} \right)+\\
    -(2x+3a_j+a_i)\ln\left( 1+\frac{x}{a_i}\right)\bigg].
    \end{aligned}
\end{equation}
Here $a_j\equiv M_j^2/M_1^2$ and $1/D_i(x)\equiv (x-a_i)/[(x-a_i)^2+a_ic_i]$ is the off-shell part of the $N_i$ propagator with $c_i=a_i(Y^\dagger Y)^2_{ii}/(8\pi)^2$.
\\
On the other hand, for the $\Delta L=1$ scattering processes, it is convenient to write the rates $S_{s,t}(z)$ in term of the functions $f_{s,t}(z)$ defined as \cite{Buchmuller:2002rq}:
\begin{equation}
    f_{s,t}(z)=\frac{\int_{z^2}^\infty d\Psi\chi_{s,t}(\Psi/z^2)\sqrt{\Psi}\mathcal{K}_1(\sqrt{\Psi})}{z^2\mathcal{K}_2(z)}\,,
\end{equation}
with the functions $\chi_{s,t}(x)$ as follows:
\begin{equation}
    \chi_s(x)=\left( \frac{x-1}{x}\right)^2,
\end{equation}
\begin{equation}
    \chi_t(x)=\frac{x-1}{x}\left[ \frac{x-2+2a_h}{x-1+a_h}+\frac{1-2a_h}{x-1}\log\left( \frac{x-1+a_h}{a_h}\right)\right]\,,
\end{equation}
where $a_h=m_\phi/M_1$ is the infrared cut-off for the t-channel and $m_\phi$ is the mass of the Higgs boson which receives contributions from the thermal bath and its value can be written as $m_\phi\simeq 0.4 \;T$ \cite{2004}. 
In this a way, the $\Delta L=1$ scattering terms are:
\begin{equation}
    S_{s,t}(z)=\frac{K_s}{9\zeta(3)}f_{s,t}(z)\,,
\end{equation}
with $K_s=\widetilde{m}_1/m_\ast^s$ and \cite{Buchmuller:2004nz}:
\begin{equation}
    m_\ast^s=\frac{4\pi^2}{9}\frac{g_N}{h_t^2}m_\ast\, ,
\end{equation}
where $h_t$ is the top quark Yukawa coupling evaluated at $T\simeq M_1$.
\\
The total scattering term $S(z)$, related to the production of the sterile neutrinos, is given by $S(z)=2S_s(z)+4S_t(z)$, where the s-channel acquires a factor of 2 stemming from the diagram involving the antiparticles, and an additional factor of 2 in the t-channel comes from the u-channel diagram. \\
Therefore, the parameter $S(z)$ can finally be written as:
\begin{equation}
    S(z)=\frac{2K_s}{9\zeta(3)}(f_s(z)+2f_t(z))\,.
\end{equation}
Thus, for the total washout term $W(z)$, which includes contributions from the inverse decay processes and the $\Delta L=1,2$ scatterings, we have \cite{2009}, \cite{Buchmuller:2002rq}:
    \begin{equation}
			    W(z)=W_{ID}(z)\left[ 1+\frac{2}{D(z)}\left( 2S_{t}(z)+\frac{N_1(z)}{N_1^{eq}(z)}S_{s}(z)+2S_N(z)+2S_{Nt}(z)\right)\right].
    \end{equation}
\bibliography{biblio}
\end{document}